\documentclass[conference]{IEEEtran}

\usepackage{cite}
\usepackage{amsmath,amssymb,amsfonts}
\usepackage{algorithm}
\usepackage{algorithmic}
\usepackage{graphicx}
\usepackage{textcomp}
\usepackage{xcolor}
\usepackage[hyphens]{url}

\usepackage{booktabs}
\usepackage{multirow}
\usepackage{enumitem}

\def\BibTeX{{\rm B\kern-.05em{\sc i\kern-.025em b}\kern-.08em
    T\kern-.1667em\lower.7ex\hbox{E}\kern-.125emX}}

\newcommand{\sysname}{AgentDSE}

\usepackage[hidelinks]{hyperref}
\newcommand\blfootnote[1]{%
  \begingroup
  \renewcommand\thefootnote{}%
  \hypersetup{hidelinks}%
  \footnotetext{#1}%
  \endgroup
}

\title{AgentDSE: Reasoning-Augmented\\Architectural Design Space Exploration}

\author{
\IEEEauthorblockN{
Chenyu Wang\textsuperscript{1,*},
Jiahe Caroline Shi\textsuperscript{2,*},
David Kong\textsuperscript{1},
Duane S. Boning\textsuperscript{2},\\
Zishen Wan\textsuperscript{1,3,$\dagger$},
Yilun Du\textsuperscript{1},
Vijay Janapa Reddi\textsuperscript{1}
}
\IEEEauthorblockA{
\textsuperscript{1}Harvard University \quad
\textsuperscript{2}Massachusetts Institute of Technology \quad
\textsuperscript{3}Columbia University
}
}

\begin{document}
\maketitle
\thispagestyle{plain}
\pagestyle{plain}
\blfootnote{$^{*}$Equal contribution.}
\blfootnote{$^{\dagger}$Corresponding author: \mbox{zishenwan@seas.harvard.edu}}

\begin{abstract}

Traditional architectural design space exploration (DSE) is highly inefficient, typically requiring tens of thousands of simulator evaluations across various optimization methods. This inefficiency arises because conventional methods treat the simulator as a black-box oracle. In contrast, human architects effectively guide exploration by reasoning through physical constraints, performance bottlenecks, data reuse, and workload structures. To bridge this gap, we introduce AgentDSE, a simulator-in-the-loop methodology driven by a general-purpose large language model (LLM) coding agent. AgentDSE automates this architectural-reasoning loop without requiring model fine-tuning, precomputed design databases, or domain-specific optimizer code. Across deep neural network (DNN) accelerator mapping, hardware/software co-design, and CPU cache-hierarchy optimization, \sysname{} achieves competitive or better design quality with up to two orders of magnitude fewer evaluations.
\sysname{} also produces inspectable traces that surface architectural hypotheses, performance cliffs, implicit priors, and simulator artifacts, making every search decision traceable rather than buried in optimizer state.


\end{abstract}

\section{Introduction}
\label{sec:intro}

\begin{figure}[t]
    \centering
    \includegraphics[width=0.24\textwidth]{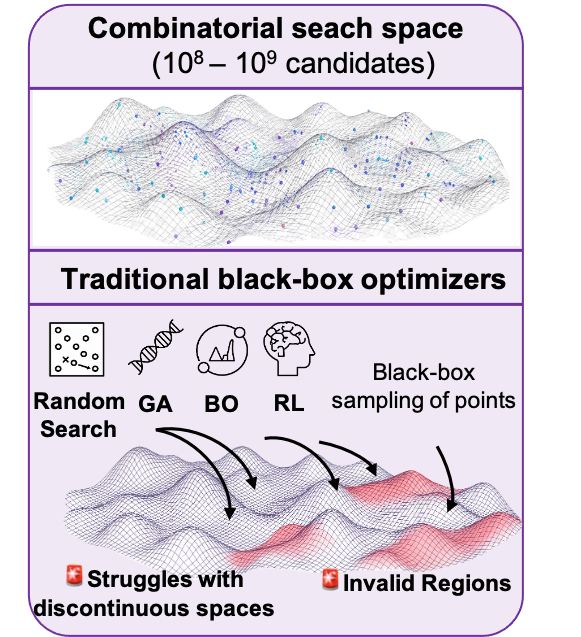}
    \hfill
    \includegraphics[width=0.24\textwidth]{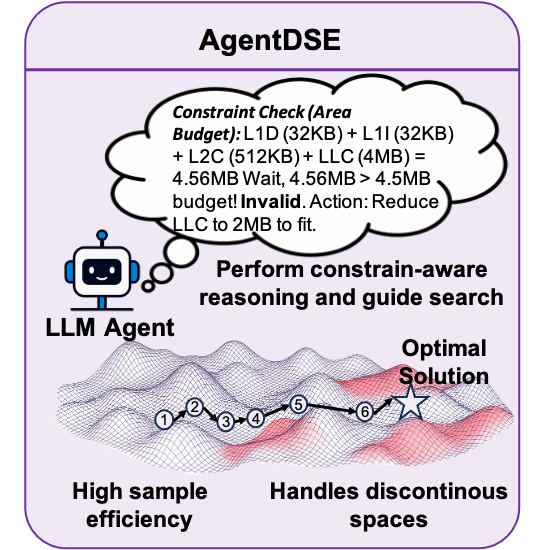}
    \caption{
Motivation for architectural-reasoning-guided DSE.
}
    \label{fig:motivation}
    \vspace{-15pt}
\end{figure}

Modern architectural design space exploration (DSE) operates over vast, discrete, and highly discontinuous spaces.
A single DNN accelerator mapping may require choosing dataflows, tiling factors, parallelization dimensions, and processing-element allocations from a combinatorial space that can exceed $10^{20}$ configurations~\cite{kwon2019maestro,kao2020gamma,krishnan2022automatic,raj2025scale}.
CPU cache-hierarchy optimization and accelerator hardware/software co-design exhibit similar structure, where each design point is constrained by hardware resources, workload behavior, and simulator-specific validity conditions.
Traditional DSE methods, including random search (RS), genetic algorithms (GA)~\cite{sun2022gibbon, kao2020gamma}, Bayesian optimization (BO)~\cite{reagen2017bo,dinh2023sample,snoek2012practical}, reinforcement learning (RL)~\cite{archgym}, and differentiable DSE methods~\cite{hong2023dosa}, introduce increasingly sophisticated samplers or surrogates over numeric design spaces.
Yet these methods largely preserve the same black-box optimization paradigm: the simulator is queried for scalar feedback, while semantic information about bottlenecks, reuse, hierarchy, invalid regions, and workload structure remains outside the search loop.
Moreover, black-box optimizers offer no insight into why a configuration was chosen, making it difficult to debug simulator artifacts or validate search decisions.
Recent work has begun exploring large language models (LLMs) for hardware design and architecture exploration~\cite{gem5copilot,llmdse,gupta2026archagent,alphazero-moment,tschand2026genai,wu2024chateda,liu2024chipnemo,ghose2025orfsagent,wang2025aiagenticprogrammingsurvey,qi2026economy}.
These efforts highlight the promise of LLM-assisted architectural optimization, but many remain tied to target-specific workflows, precomputed databases, custom multi-agent systems, or specialized prompting pipelines.
What is still missing is a reusable mechanism for turning this emerging reasoning capability into simulator-call selection in DSE.
In human architectural practice, such reasoning does not take the form of black-box sampling. Architects use a small number of observations to infer constraints, bottlenecks, data reuse, parallelism, hierarchy, and workload structure, and then decide which configurations are worth evaluating next.
Figure~\ref{fig:motivation} illustrates this contrast.

In this work, we bridge this gap by proposing \sysname{}, a simulator-in-the-loop agentic framework that puts architectural reasoning in the optimization loop.
Instead of treating DSE as purely numerical sampling, we formulate it as an iterative, history-aware reasoning process driven by an LLM agent.
Operating over a structured workspace of prior observations and hypotheses, the agent interprets task briefs through an autonomous hypothesis-test-refine loop: it proposes candidates, parses simulator feedback, and refines its hypotheses.
This approach couples the architectural knowledge of foundation models~\cite{tschand2026genai,prakash2025quarch} with a direct, grounded optimization loop.



\begin{figure*}[t]
    \centering
    \includegraphics[width=1\textwidth]{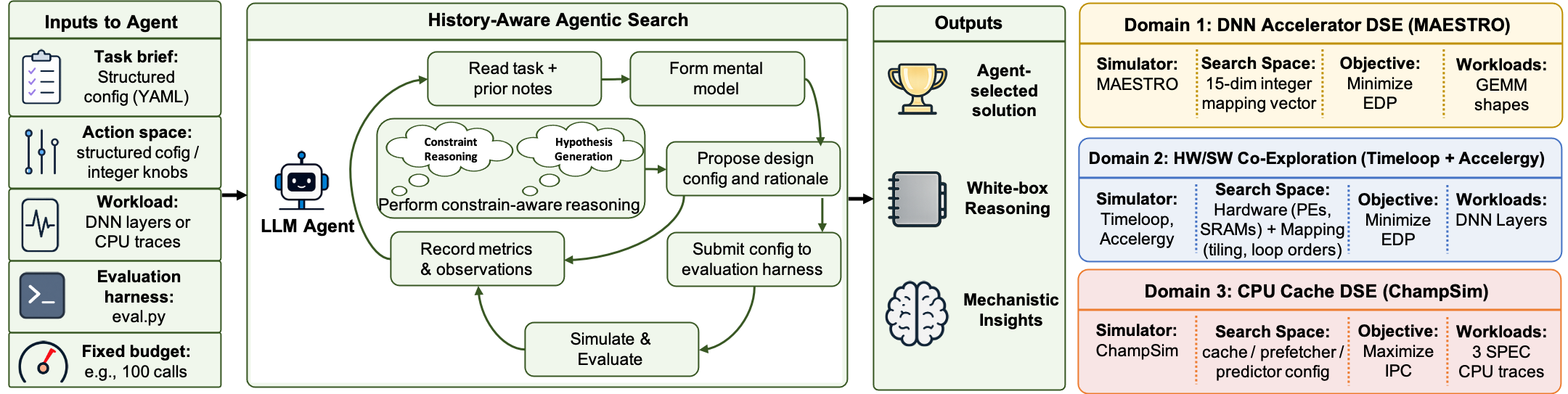}
    \caption{
    Overview of \sysname{} for history-aware agentic design-space search.
    }
    \label{fig:framework}
\end{figure*}

We make the following contributions:
\begin{itemize}[leftmargin=*,itemsep=2pt]
    \item \textbf{An Agentic DSE Paradigm:} We formalize \sysname{} as a domain-agnostic agent-simulator workflow that runs architectural reasoning as an iterative hypothesis-test-refine loop driven by simulator feedback. The same workspace abstraction is reused across MAESTRO, Timeloop/Accelergy, and ChampSim with no fine-tuning and no precomputed design database.
    \item \textbf{Cross-domain Sample Efficiency Gains:} We evaluate \sysname{} across DNN accelerator mapping, hardware/software co-design, and CPU cache-hierarchy optimization. Across these settings, \sysname{} achieves competitive or better design quality with up to two orders of magnitude fewer simulator evaluations.
    We further decompose this gain through a controlled anonymization ablation on a representative ResNet50 layer, separating the LLM's strength as a black-box combinatorial optimizer from the additional acceleration provided by domain-specific architectural reasoning.
    \item \textbf{Auditable Optimization Traces:} 
    \sysname{} produces inspectable hypothesis-test-refine traces that surface architectural insights, biased priors, non-monotonic performance cliffs, and simulator artifacts, making every search decision traceable and
verifiable.
\end{itemize}



\section{Agentic DSE Methodology}
\label{sec:method}

Figure~\ref{fig:framework} shows the \sysname{} workflow. \sysname{} casts
design-space exploration as a simulator-in-the-loop coding task. Each run is
packaged as a self-contained workspace that specifies the design objective,
candidate schema, workload, hardware constraints, evaluator, and evaluation
budget. A coding agent then iteratively edits candidate designs, invokes the
evaluator, observes measured feedback, and records search state in workspace
artifacts.

The methodological move is not simply attaching an LLM to a simulator, but
exposing DSE as a persistent, editable, semantically grounded workspace in
which the agent accumulates hypotheses, tests them against simulator feedback,
and revises subsequent actions, shifting what the optimizer reads and
writes from a scalar reward signal to a structured, inspectable record.

\subsection{Workspace Interface}

An \sysname{} workspace contains the following artifacts.
\begin{itemize}[leftmargin=*,itemsep=1pt]
    \item \textbf{Task brief} (\texttt{README.md}): describes the objective,
    domain context, budget, and expected workflow.
    \item \textbf{Action-space specification} (\texttt{action\_space.md}):
    defines candidate fields, valid ranges, and constraints.
    \item \textbf{Workload and hardware files} (\texttt{workload.yaml},
    \texttt{hardware.yaml}): specify the target workload and either fixed
    hardware parameters or hardware search-space bounds.
    \item \textbf{Candidate file} (\texttt{candidate.yaml}): stores the next
    design to evaluate.
    \item \textbf{Evaluation harness} (\texttt{submit.sh}, \texttt{eval.py}):
    validates the candidate, invokes the simulator, updates logs, and returns
    measured metrics.
    \item \textbf{Search state} (\texttt{notes.md}, \texttt{history.jsonl},
    \texttt{best.json}, \texttt{budget.json}): records free-form observations,
    all evaluations, the best design found so far, and remaining budget.
\end{itemize}

This interface deliberately exposes the DSE problem through ordinary files rather
than through a custom optimizer API. Porting \sysname{} to a new domain requires only a domain-specific candidate schema and an evaluation harness that maps candidates to measured metrics.

\subsection{Optimization Loop}

Algorithm~\ref{alg:loop} summarizes the optimization procedure:

\begin{algorithm}[h]
\small
\caption{\sysname{} Optimization Loop}
\label{alg:loop}
\begin{algorithmic}[1]
\STATE Read \texttt{README.md}, \texttt{action\_space.md}, \texttt{workload.yaml}
\STATE Initialize search state from \texttt{notes.md}, \texttt{history.jsonl}, and \texttt{best.json}.
\WHILE{budget remaining and agent has not terminated}
    \STATE \textbf{Reason} about the next action: form hypotheses about which design dimensions to explore and filter candidates against the constraints
    \STATE Write candidate to \texttt{candidate.yaml}
    \STATE Run \texttt{./submit.sh} $\to$ receive metrics
    \STATE Append the result to \texttt{history.jsonl} and update \texttt{best.json} if improved
    \STATE Update \texttt{notes.md} with observations and next-step rationale.
\ENDWHILE
\STATE Return the best recorded design
\end{algorithmic}
\end{algorithm}

The key distinction from prior LLM-for-DSE systems is the absence of a domain-specific control framework: no state machine~\cite{gem5copilot}, no multi-agent orchestration~\cite{llmdse}, and no retrieval over pre-computed design databases.
The agent's ``framework'' is simply its native coding loop (read, edit, and execute) applied to simulator-driven optimization.

\subsection{Auditable Optimization Traces}

\sysname{} produces more than a final design point.
Because the agent operates through editable files and explicit tool calls, the optimization process is externalized into persistent workspace artifacts.
In particular, \texttt{notes.md}, \texttt{history.jsonl}, \texttt{best.json},
and post-processed Thinking-Action-Observation (TAO) traces together form an \emph{auditable optimization trace}: they record the agent's intermediate hypotheses, the candidates it evaluated, the measured outcomes it observed, and the best-so-far progress over the course of search.

These traces are method artifacts rather than incidental logs.
They make the search process inspectable at the level of concrete design decisions and their measured consequences, enabling post hoc analysis of why a design improved, why a trajectory failed, or whether an apparent gain was caused by a simulator artifact.
As a result, \sysname{} exposes both an optimization \emph{outcome} and a traceable optimization \emph{process}.

\subsection{Reusing the Workflow Across Domains}

Figure~\ref{fig:framework} shows how the same workspace abstraction instantiates across three qualitatively different simulator stacks. Although the simulators target different design problems, they share the same optimization interface.
In MAESTRO~\cite{kwon2019maestro}, an analytical cost model for DNN accelerator mappings, the candidate file encodes mapping decisions such as loop order, tiling, spatial dimensions, and PE allocation.
In Timeloop/Accelergy~\cite{parashar2019timeloop,wu2019accelergy}, a mapper and energy-modeling stack for tiled accelerators, the candidate specifies a joint hardware/software co-design point.
In ChampSim~\cite{champsim}, a trace-driven CPU microarchitecture simulator, the candidate encodes cache-hierarchy, prefetcher, and branch-predictor choices under structural constraints.
Across all three settings, the agent edits the same candidate artifact, invokes the simulator, and records persistent search traces, allowing \sysname{} to reuse one coding-agent workflow without per-domain optimizer.

Taken together, this workflow exposes more than a scalar objective: it gives the agent semantic context, explicit constraint feedback, and persistent search history.
These properties give the agent the inputs it needs to reason about each simulator call, and may enable more structured candidate selection than purely black-box or learned optimizers.
We revisit this intuition in Section~\ref{sec:analysis} through optimization traces, case studies, and ablations.

\section{Experimental Setup}
\label{sec:setup}

To demonstrate the effectiveness of \sysname{}, we conduct three studies. First, we perform a head-to-head comparison with DOSA~\cite{hong2023dosa} by fully replicating their experimental setup and substituting their DSE engine with \sysname{} under identical constraints. Second, we evaluate the generalization and scalability of our approach on MAESTRO~\cite{kwon2019maestro} and ChampSim~\cite{champsim} simulators, comparing against traditional baselines including Random Search (RS), Genetic Algorithms (GA), and Reinforcement Learning (PPO). Finally, we conduct an ablation that separates \sysname{}'s architectural reasoning from its combinatorial optimization capability to better understand the underlying drivers of its performance.

These three thrusts play complementary evaluative roles: the DOSA setting provides a direct matched-setup comparison against a representative differentiable DSE engine; MAESTRO tests whether the same agentic workflow transfers to a different DNN-mapping problem under a strict simulator-call budget against classical baselines; and ChampSim probes cross-domain transfer to a CPU memory-hierarchy task and the convergence efficiency of the agent's self-terminating search. Each setting emphasizes a different evaluation question, so the baselines and metrics are chosen to match established practice in that domain.

\subsection{Primary Setting: Co-Design with DOSA/Timeloop}
\label{sec:setup-dosa}

For our primary evaluation, we establish a head-to-head comparison against DOSA~\cite{hong2023dosa}, a representative differentiable DSE methodology for accelerator co-design.
The setting is a hardware/software co-design problem targeting a Gemmini-style systolic array~\cite{genc2021gemmini} using the Timeloop/Accelergy evaluation stack~\cite{parashar2019timeloop,wu2019accelergy}.
The optimizer must jointly select architectural parameters (e.g., spatial dimensions, buffer sizes) and dataflow mappings.
While DOSA leverages a pre-computed dataset of 72{,}701 Timeloop samples (covering 65 distinct workload shapes) and performs 1{,}490 gradient-descent steps on the resulting analytical predictor across the 24-layer sweep ($\sim$62 steps/layer, with no real Timeloop calls in this inner loop), \sysname{} explores the discrete space from a cold start (no precomputed dataset and no surrogate predictor) under a strict budget of 80 ground-truth Timeloop calls per layer. We evaluate on the same 24 representative ResNet50 layers used in the DOSA study, targeting Energy-Delay Product (EDP). Detailed cost accounting and the inclusive per-layer comparison are deferred to Section~\ref{sec:res-dosa}.

\subsection{Generalization: MAESTRO and ChampSim}

To evaluate the generalization and scalability of our approach, we deploy the identical agentic methodology on two additional DSE domains.

\textbf{DNN Accelerator Mapping (MAESTRO).} 
We optimize spatial mapping and tiling on an edge-constrained accelerator using MAESTRO~\cite{kwon2019maestro}. The action space spans 15 integer dimensions. We target EDP minimization across 10 distinct general matrix multiply (GEMM) shapes drawn from LLaMA-3.1-8B~\cite{llama3} and OpenVLA~\cite{openvla}. \sysname{} is restricted to 100 simulator calls per shape, compared against RS and GA at 20,000 calls, and Proximal Policy Optimization (PPO) at 50,000 steps.

\textbf{CPU Cache-Hierarchy (ChampSim).}
We optimize a CPU memory hierarchy (L1/L2/last-level cache (LLC) sizes, associativities, prefetchers, and replacement policies) subject to a strict 4.5MiB static RAM (SRAM) area constraint using ChampSim~\cite{champsim}. We target Instructions Per Cycle (IPC) maximization on three distinct SPEC CPU2017 traces, with three seeds per workload (9 runs total). \sysname{} autonomously terminates after $\sim$50 calls, while the baselines of RS, GA, and BO are assigned 1,000 evaluations.
\section{Results}
\label{sec:results}

\subsection{Head-to-Head: Differentiable DSE vs. \sysname{}}
\label{sec:res-dosa}

\begin{figure}[t]
    \centering
    \includegraphics[width=\columnwidth]{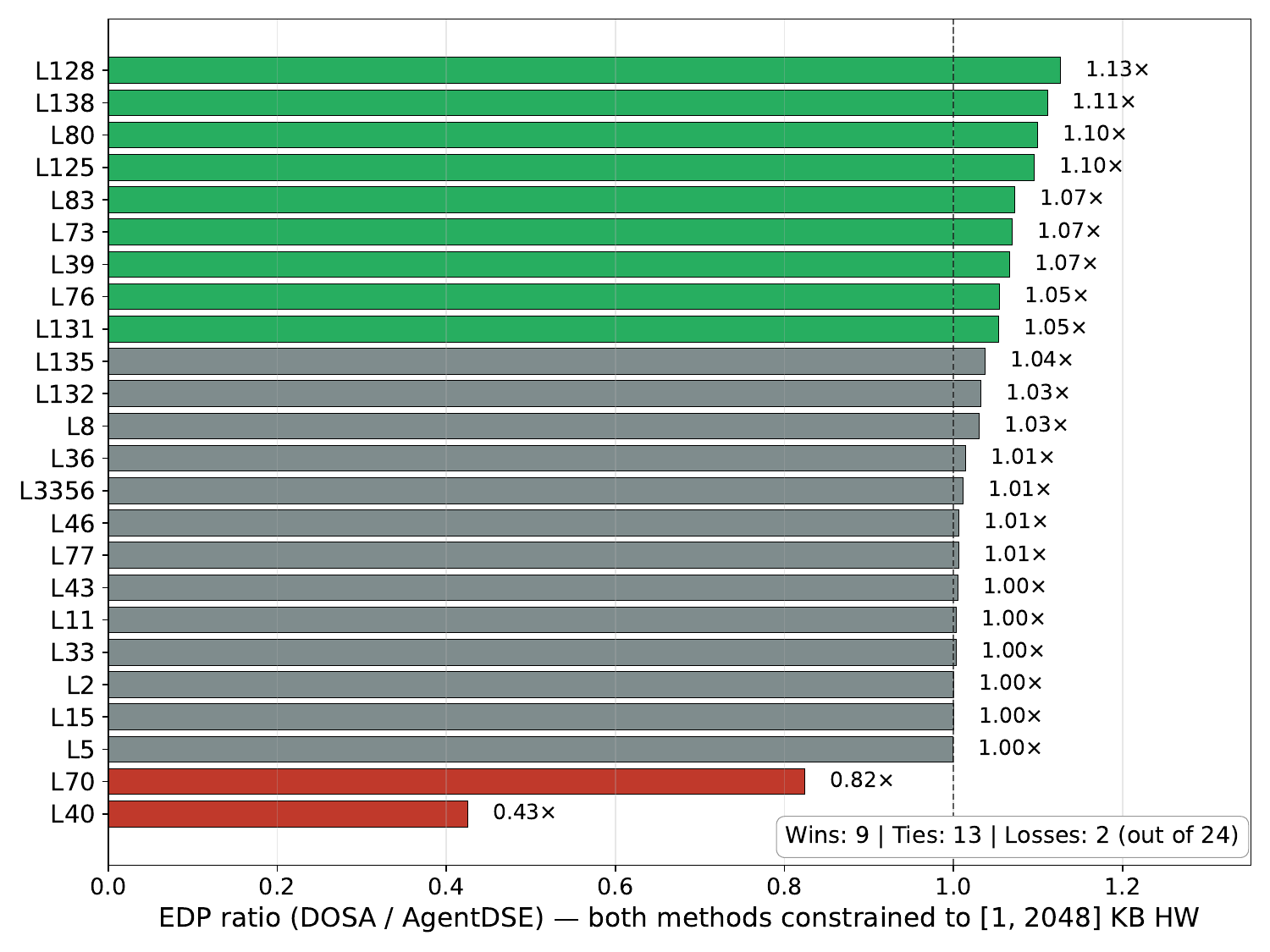}
    \caption{Strict apples-to-apples per-layer comparison: DOSA in per-layer
    mode with HW search bounds matched to \sysname{}'s $[1, 2048]$~KB via a
    patched gradient-descent penalty. $9$ wins, $13$ ties, $2$ losses;
    geometric-mean EDP ratio $0.99\times$.}
    \label{fig:gemmini-perlayer-bounded}
\end{figure}

We compare \sysname{} against DOSA under two strict apples-to-apples scopes that hold optimization scope fixed between the two methods: a per-layer comparison (Table~\ref{tab:gemmini-perlayer}, with Figure~\ref{fig:gemmini-perlayer-bounded} as visual companion) where each method optimizes each ResNet50 layer independently, and a network-level comparison (Table~\ref{tab:network-headtohead}) where each method optimizes one shared HW configuration across all 24 layers. Across both scopes, \sysname{} matches DOSA's EDP.
For the per-layer comparison, we use a $\pm 5\%$ equivalence band to avoid over-interpreting small numerical EDP differences. \sysname{} achieves $9$ wins, $13$ ties, and $2$ losses, with a geometric-mean EDP ratio of $0.99\times$. At the network level, \sysname{} achieves a $1.15\times$ improvement while using an average of $31$ ground-truth simulator calls per layer, a $\sim$$37\times$ reduction relative to DOSA's inclusive cost.

\textbf{Cost accounting for DOSA.}
DOSA's reported cost has three components: (i) a 72{,}701-sample offline Timeloop dataset covering 65 distinct workload shapes, used to train an analytical predictor; (ii) for each layer, 7 random initializations of gradient descent on that predictor, each running 1{,}490 steps ($7 \times 1{,}490 = 10{,}430$ predictor inferences per layer, with no real Timeloop calls in this inner loop); and (iii) 7 real Timeloop calls per layer at search time to validate the predicted optima. Amortizing the 72{,}701-sample offline dataset across the 65 shapes it covers yields $\sim$1{,}118 Timeloop calls per shape; each ResNet50 layer maps to one of these covered shapes (the dataset is constructed to include the layer dimensions used in DOSA's evaluation), so adding the 7 search-time calls per layer gives an inclusive cost of $\sim$1{,}125 real Timeloop calls per layer. Under this inclusive accounting, \sysname{}'s cold-start search uses $\sim$\textbf{37$\times$} fewer real Timeloop calls per layer than DOSA. We adopt inclusive accounting as the fairer common cost basis because \sysname{} has no comparable offline sampling phase to amortize: it issues real Timeloop calls from a cold start, with no precomputed dataset and no surrogate predictor. A search-only view, which charges DOSA only its 7 post-predictor verification calls per layer, omits the 72{,}701-sample modeling investment that any new accelerator or simulator stack would have to repeat.

\begin{table}[h]
\small
\centering
\caption{\textbf{Strict per-layer head-to-head: \sysname{} matches DOSA's per-layer EDP (9 wins, 13 ties, 2 losses; geometric-mean ratio $0.99\times$) at $\sim$$37\times$ lower sample cost.} Bold marks the row winner; ties (within $5\%$) are bolded by lower EDP. Protocol details and cost accounting are in Section~\ref{sec:res-dosa}.}
\label{tab:gemmini-perlayer}
\setlength\tabcolsep{3pt}
\resizebox{\linewidth}{!}{%
\begin{tabular}{@{}llccc@{}}
\toprule
\textbf{Layer} & \textbf{Dimensions $(K, C, H, W, R, S)$} & \textbf{\sysname{}} & \textbf{DOSA} & \textbf{EDP Ratio} \\
 & & (J$\cdot$cyc) & (J$\cdot$cyc) & DOSA/\sysname{} \\
\midrule
Conv83 & $256 {\times} 256 {\times} 14 {\times} 14 {\times} 3 {\times} 3$ & \textbf{19.0} & 20.4 & 1.07$\times$ \\
Conv76 & $1024 {\times} 256 {\times} 14 {\times} 14 {\times} 1 {\times} 1$ & \textbf{7.01} & 7.39 & 1.05$\times$ \\
Conv8 & $64 {\times} 64 {\times} 56 {\times} 56 {\times} 3 {\times} 3$ & \textbf{10.1} & 10.4 & 1.03$\times$ \\
Conv39 & $512 {\times} 128 {\times} 28 {\times} 28 {\times} 1 {\times} 1$ & \textbf{8.11} & 8.64 & 1.07$\times$ \\
Conv46 & $128 {\times} 128 {\times} 28 {\times} 28 {\times} 3 {\times} 3$ & \textbf{7.70} & 7.75 & 1.01$\times$ \\
Conv131 & $2048 {\times} 512 {\times} 7 {\times} 7 {\times} 1 {\times} 1$ & \textbf{26.3} & 27.7 & 1.05$\times$ \\
Conv2 & $64 {\times} 3 {\times} 112 {\times} 112 {\times} 7 {\times} 7$ & \textbf{235} & 235 & 1.00$\times$ \\
Conv5 & $64 {\times} 64 {\times} 56 {\times} 56 {\times} 1 {\times} 1$ & 2.79 & \textbf{2.79} & 1.00$\times$ \\
Conv11 & $256 {\times} 64 {\times} 56 {\times} 56 {\times} 1 {\times} 1$ & \textbf{20.4} & 20.5 & 1.00$\times$ \\
Conv15 & $64 {\times} 256 {\times} 56 {\times} 56 {\times} 1 {\times} 1$ & 20.5 & 20.5 & 1.00$\times$ \\
Conv33 & $128 {\times} 256 {\times} 56 {\times} 56 {\times} 1 {\times} 1$ & \textbf{37.1} & 37.2 & 1.00$\times$ \\
Conv36 & $128 {\times} 128 {\times} 28 {\times} 28 {\times} 3 {\times} 3$ & \textbf{17.0} & 17.3 & 1.01$\times$ \\
Conv40 & $512 {\times} 256 {\times} 28 {\times} 28 {\times} 1 {\times} 1$ & 88.0 & \textbf{37.4} & 0.43$\times$ \\
Conv43 & $128 {\times} 512 {\times} 28 {\times} 28 {\times} 1 {\times} 1$ & \textbf{8.30} & 8.34 & 1.00$\times$ \\
Conv70 & $256 {\times} 512 {\times} 28 {\times} 28 {\times} 1 {\times} 1$ & 22.1 & \textbf{18.2} & 0.82$\times$ \\
Conv73 & $256 {\times} 256 {\times} 14 {\times} 14 {\times} 3 {\times} 3$ & \textbf{25.7} & 27.5 & 1.07$\times$ \\
Conv77 & $1024 {\times} 512 {\times} 14 {\times} 14 {\times} 1 {\times} 1$ & \textbf{33.8} & 34.0 & 1.01$\times$ \\
Conv80 & $256 {\times} 1024 {\times} 14 {\times} 14 {\times} 1 {\times} 1$ & \textbf{7.48} & 8.22 & 1.10$\times$ \\
Conv125 & $512 {\times} 1024 {\times} 14 {\times} 14 {\times} 1 {\times} 1$ & \textbf{21.8} & 23.9 & 1.10$\times$ \\
Conv128 & $512 {\times} 512 {\times} 7 {\times} 7 {\times} 3 {\times} 3$ & \textbf{133} & 149 & 1.13$\times$ \\
Conv132 & $2048 {\times} 1024 {\times} 7 {\times} 7 {\times} 1 {\times} 1$ & \textbf{112} & 116 & 1.03$\times$ \\
Conv135 & $512 {\times} 2048 {\times} 7 {\times} 7 {\times} 1 {\times} 1$ & \textbf{28.0} & 29.0 & 1.04$\times$ \\
Conv138 & $512 {\times} 512 {\times} 7 {\times} 7 {\times} 3 {\times} 3$ & \textbf{125} & 139 & 1.11$\times$ \\
FC\_out & $1000 {\times} 2048 {\times} 1 {\times} 1 {\times} 1 {\times} 1$ & \textbf{54.2} & 54.8 & 1.01$\times$ \\
\bottomrule
\end{tabular}
}
\end{table}

\paragraph{Per-layer protocol and takeaways}
The DOSA column in Table~\ref{tab:gemmini-perlayer} comes from DOSA in
per-layer mode: single-layer workloads, $\textsc{num\_starts}{=}7$ random
initializations per layer, with the gradient-descent loop patched with a
steep quadratic-plus-cubic HW-bound penalty plus a hard projection step so
DOSA's trajectory is constrained to \sysname{}'s $[1, 2048]$~KB
scratchpad/accumulator search bounds. \sysname{} wins on $9$ layers, ties on
$13$, and loses on $2$ across 24 layers, with a geometric-mean EDP ratio of
$0.99\times$ (median $1.02\times$). The two losses (Conv40 at $0.43\times$
and Conv70 at $0.82\times$) are pointwise wins for DOSA on individual layers
but do not propagate to the network level as shown in Table~\ref{tab:network-headtohead},
where \sysname{} wins outright.
The headline takeaway at this protocol is that \sysname{} reaches
DOSA-comparable per-layer EDP \emph{without} DOSA's 72{,}701-sample offline
predictor training set; the dominant remaining advantage is on sample
efficiency rather than EDP itself.

\paragraph{Network-level head-to-head}
At the network level (one shared HW driving all 24 layers, the same setting
as DOSA's published recipe), \sysname{} achieves a mean network EDP of
$6.62 \times 10^{10}$~uJ$\cdot$cyc averaged across three random seeds,
compared to DOSA's reproduced best of $7.59 \times 10^{10}$
($\mathbf{1.15\times}$ \sysname{} improvement) and DOSA's paper-reported
$\sim$$1.0 \times 10^{11}$. All three \sysname{} seeds converged to the same
shared HW configuration (pe\_dim$=128$, sp\_size$=512$~KB,
acc\_size$=512$~KB), so the cross-seed variance enters only through the
per-layer mapping search and the number of evaluations to convergence;
seeds used $\sim$$18.5$ budget units on average (each unit invokes Timeloop
on all 24 layers, so $\sim$$444$ underlying Timeloop calls per seed) at a
mean API cost of $\sim$\$12. While DOSA's gradient descent must trade off
HW size against per-layer fitness through a smooth predictor loss
landscape, \sysname{} explicitly chose the maximum allowed PE dimension and
balanced buffer sizes by reasoning over per-layer mapping requirements ---
a discrete design move that gradient descent does not naturally discover.

\begin{table}[h]
\small
\centering
\caption{\textbf{Network-level head-to-head: \sysname{} achieves $\mathbf{1.15\times}$ lower network EDP than DOSA's reproduction with $\sim$$165\times$ fewer real Timeloop calls.} \sysname{} values are mean across 3 random seeds; cost-accounting breakdown for DOSA's $72{,}869$ inclusive calls is given above.}
\label{tab:network-headtohead}
\setlength\tabcolsep{4pt}
\resizebox{\linewidth}{!}{%
\begin{tabular}{@{}lccc@{}}
\toprule
\textbf{Method} & \textbf{Network EDP} & \textbf{Real Timeloop calls} & \textbf{API Cost} \\
 & ($\mu$J$\cdot$cyc) & & (USD) \\
\midrule
DOSA (estimated from paper)      & $\sim$$1.0 \times 10^{11}$ & $72{,}869$ & --- \\
DOSA (our reproduction, best of 3) & $7.59 \times 10^{10}$    & $72{,}869$ & --- \\
\sysname{} (network, mean of 3 seeds) & $\mathbf{6.62 \times 10^{10}}$ & $\mathbf{\sim 444}$ & $\mathbf{\sim\$12}$ \\
\midrule
\textbf{\sysname{} vs.\ DOSA reproduction} & \textbf{$1.15\times$ lower EDP} & \textbf{$\sim$$165\times$ fewer Timeloop calls} & --- \\
\bottomrule
\end{tabular}
}
\end{table}

\sysname{} also self-terminates well below its 50-budget-unit ceiling,
indicating its convergence detection is not an artifact of budget.

\subsection{Anonymization Ablation: Reasoning vs.\ Optimization}
\label{sec:res-ablation}

To isolate the source of this efficacy, we conduct a controlled anonymization ablation on Layer 46. By stripping all semantic meaning from the workspace (i.e., hiding parameter names and physical metrics), we force the agent to optimize a purely numerical integer space.

\begin{table}[h]
\small
\centering
\caption{\textbf{Anonymization ablation on ResNet50 Layer~46: both agent variants tie DOSA's strict per-layer EDP; restoring physical context cuts calls from 71 to 29 (a $\sim$$2.4\times$ sample-efficiency speedup).} The \emph{Anonymized Agent} sees only opaque integer IDs; the \emph{Full-Context Agent} sees physical names and units. DOSA baseline as in Table~\ref{tab:gemmini-perlayer}; cost-accounting in Section~\ref{sec:res-dosa}.}
\label{tab:ablation}
\resizebox{\linewidth}{!}{%
\begin{tabular}{@{}lcccc@{}}
\toprule
\textbf{Optimizer Setup} & \textbf{Best EDP (J$\cdot$cyc)} & \textbf{EDP ratio} & \textbf{Timeloop calls} & \textbf{Timeloop calls (\% of DOSA)} \\
 & & DOSA $/$ Agent & & \\
\midrule
DOSA (per-layer mode) & 7.75 & 1.00$\times$ & $\sim$1{,}125 & 100\% \\
\midrule
Anonymized Agent & 7.79 & 0.99$\times$ & 71 & 6.3\% \\
Full-Context Agent & \textbf{7.70} & \textbf{1.01$\times$} & \textbf{29} & \textbf{2.6\%}\\
\bottomrule
\end{tabular}%
}
\end{table}

\begin{figure}[t]
    \centering
    \includegraphics[width=\columnwidth]{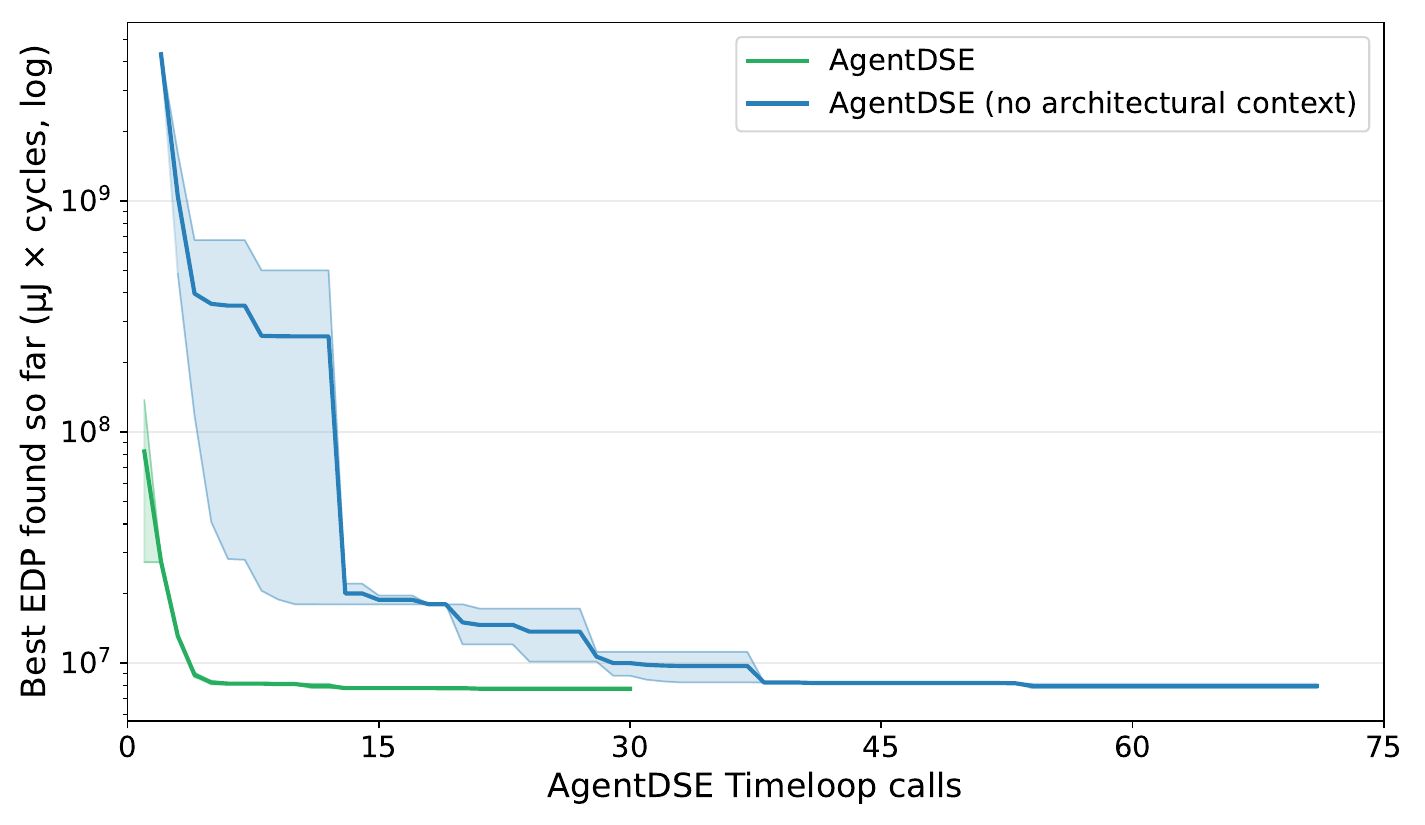}
    \caption{Convergence on ResNet50 Layer~46: best-so-far EDP versus real Timeloop calls
    for the Full-Context and Anonymized \sysname{} variants (29 and 71 calls to converge,
    respectively). DOSA's $\sim$$1{,}125$-call inclusive per-layer budget does not fit
    visually on this axis; its EDP comparison is in
    Figure~\ref{fig:gemmini-perlayer-bounded} and
    Tables~\ref{tab:gemmini-perlayer},~\ref{tab:network-headtohead}.}
    \label{fig:convergence-l46}
\end{figure}

The results (Table~\ref{tab:ablation}) point to two effects.
First, the \emph{Anonymized Agent} reaches DOSA-level per-layer EDP on this layer (within $1\%$) using only $71$ Timeloop calls instead of $\sim$1{,}125, indicating that the LLM is itself a capable black-box combinatorial optimizer at vastly lower sample cost, even with no architectural semantics.
Restoring the physical semantics (\emph{Full-Context Agent}) further reduces the simulator calls required from 71 to 29 (a $\sim$2.4$\times$ speedup) while reaching the same EDP quality.
On this layer, then, the LLM-as-optimizer drives the headline sample-efficiency win, while architectural reasoning provides a multiplier on top of it; the two effects are complementary rather than substitutive.
Figure~\ref{fig:convergence-l46} visualizes the corresponding convergence trajectories: both agent variants reach DOSA-comparable EDP within tens of simulator calls, with the full-context configuration converging earliest.

\subsection{Cross-Domain Generalization}

We summarize \sysname{}'s performance on the remaining two hardware domains to test whether this sample efficiency generalizes.

\textbf{MAESTRO DNN Mapping.}
As shown in Table~\ref{tab:model-edp}, \sysname{} achieves the best model-level EDP on both LLaMA-3.1-8B and OpenVLA while using only 100 simulator calls per shape. In contrast, RS and GA use 20,000 evaluations per shape, and PPO uses 50,000 evaluations. Thus, \sysname{} attains the best EDP among the four methods evaluated while using a 200--500$\times$ smaller simulator budget.

\begin{table}[h]
\small
\caption{\textbf{\sysname{} attains the best model-level EDP on both LLaMA-3.1-8B and OpenVLA at 200--500$\times$ smaller simulator budget.} MAESTRO weighted EDP, scaled by $10^{-15}$; lower is better, \textbf{bold} marks the row winner. Per-shape simulator budgets in parentheses.}
\label{tab:model-edp}
\centering
\resizebox{\linewidth}{!}{%
\begin{tabular}{@{}lcccc@{}}
\toprule
\textbf{Model} & \textbf{\sysname{}} & \textbf{RS} & \textbf{GA} & \textbf{PPO} \\
 & \textbf{(100/shape)} & \textbf{(20K)} & \textbf{(20K)} & \textbf{(50K)} \\
\midrule
LLaMA-3.1-8B & \textbf{28.0} & 85.7 & 30.9 & 29.4 \\
OpenVLA      & \textbf{7.17} & 19.0 & 7.83 & 8.03 \\
\bottomrule
\end{tabular}
}
\end{table}

These aggregate results do not imply exhaustive coverage of every per-layer
optimum; rather, they show that reasoning-guided search can identify strong
model-level designs under a much smaller simulator budget.

\textbf{ChampSim CPU Cache Hierarchy.}
We report ChampSim along three views, since they answer different questions: (i) \emph{final IPC} at each method's own budget (Table~\ref{tab:champsim-per-workload}); (ii) \emph{equal-quality match-point}---the smallest \sysname{} budget that first reaches Random's 1{,}000-call IPC (Table~\ref{tab:champsim-budget-scan}); and (iii) \emph{convergence-matched cost}---calls until each method plateaus (Table~\ref{tab:champsim-convergence}). Final IPC reaches parity (within $\sim1.5\%$, bounded by simulator variance limits) with traditional optimizers given 1{,}000 evaluations, while \sysname{} self-terminates after $\sim$50 calls; the equal-quality and convergence-matched views then make the sample-efficiency advantage explicit.

\begin{table}[h]
\small
\caption{\textbf{\sysname{} matches RS, GA, and BO on final IPC while using $\sim$20$\times$ fewer simulator calls.} Best IPC per ChampSim workload (higher is better, \textbf{bold} = row winner). \sysname{}: $\sim$50 self-terminated calls; RS/GA/BO: 1{,}000 calls each. On \texttt{lbm} and \texttt{omnetpp} all four methods cluster within $\sim$1.5\% of the row winner, well within the 5\,M-instruction simulator's run-to-run variance, so we treat these rows as ties.}
\label{tab:champsim-per-workload}
\centering
\resizebox{\linewidth}{!}{%
\begin{tabular}{@{}lcccc@{}}
\toprule
\textbf{Workload} & \textbf{\sysname{}} & \textbf{Random} & \textbf{GA} & \textbf{BO} \\
 & \textbf{($\sim$50 calls)} & \textbf{(1000)} & \textbf{(1000)} & \textbf{(1000)} \\
\midrule
\texttt{mcf}     & \textbf{0.6141} & 0.5941 & 0.6120          & 0.6112 \\
\texttt{lbm}     & 0.5771          & 0.5752 & 0.5854          & \textbf{0.5858} \\
\texttt{omnetpp} & 0.3166          & 0.3176 & \textbf{0.3193} & 0.3181 \\
\bottomrule
\end{tabular}
}
\end{table}

\begin{table}[h]
\small
\caption{\textbf{At equal IPC, \sysname{} reaches Random's 1{,}000-call quality in 20--50 ChampSim calls (20--50$\times$ speedup).} ``\sysname{} ChampSim calls'' is the earliest checkpoint matching Random@1000; speedup is $1000\,/\,\text{calls}$. On \texttt{omnetpp} all four methods cluster within 0.002 IPC at every budget, so a match point cannot be defined.}
\label{tab:champsim-budget-scan}
\centering
\resizebox{\linewidth}{!}{%
\begin{tabular}{@{}lccc@{}}
\toprule
\textbf{Workload} & \textbf{Random@1000 IPC} & \textbf{\sysname{} ChampSim calls} & \textbf{Speedup} \\
\midrule
\texttt{mcf}     & 0.594 & \textbf{20}  & 50$\times$ \\
\texttt{lbm}     & 0.575 & \textbf{50}  & 20$\times$ \\
\texttt{omnetpp} & 0.318 & no match point & --- \\
\bottomrule
\end{tabular}
}
\end{table}

To provide a tighter comparison, we evaluate convergence-matched sample efficiency. Table~\ref{tab:champsim-convergence} compares the number of calls required for each method to converge. Convergence-matched, \sysname{}'s sample-efficiency advantage is \textbf{7--15$\times$} across all workloads and baselines.

\begin{table}[h]
\small
\caption{\textbf{Convergence-matched, \sysname{} is 7--15$\times$ more sample-efficient than RS, GA, and BO on every workload.} The \sysname{} column shows calls to plateau (no IPC improvement over the next 20 calls); the others report \sysname{}'s speedup at each baseline's own convergence point.}
\label{tab:champsim-convergence}
\centering
\resizebox{\linewidth}{!}{%
\begin{tabular}{@{}lcccc@{}}
\toprule
\textbf{Workload} & \textbf{\sysname{} ChampSim calls} & \textbf{vs RS} & \textbf{vs GA} & \textbf{vs BO} \\
\midrule
\texttt{mcf}     & 55 & 8.5$\times$ & 9.9$\times$ & 11.4$\times$ \\
\texttt{lbm}     & 54 & 9.2$\times$ & 6.9$\times$ & 11.2$\times$ \\
\texttt{omnetpp} & 41 & 15.4$\times$ & 9.8$\times$ & 14.3$\times$ \\
\bottomrule
\end{tabular}
}
\end{table}



\section{Analysis and Interpretability}
\label{sec:analysis}

The sample-efficiency results in Section~\ref{sec:results} say
\emph{how well} \sysname{} performs; this section examines \emph{how}
it gets there. We present two illustrative case studies drawn from the
trace logs (Section~\ref{sec:insights}) and one structural regularity
observed across runs (Section~\ref{sec:priors}); these illustrate the
kinds of design reasoning the trace exposes rather than serve as a
statistical base.

\subsection{Illustrative Trace Case Studies}
\label{sec:insights}

\sysname{}'s \texttt{notes.md} records the hypothesis behind each
candidate and its measured outcome. We highlight two example patterns
that an architect can read directly from the trace as design rules.

\paragraph{Workload-specific design rules absent from the prompt}
On the ChampSim mcf trace (latency-bound, pointer-chasing), 
the agent forms an unusual conjecture about prefetcher utility:
\begin{quote}\small\itshape
``mcf is dominated by 200+-cycle DRAM latency. Aggressive prefetchers 
will inflate MPKI but useful prefetches still hide latency, MPKI is a 
proxy for miss rate, not for end-to-end performance.''
\end{quote}
The agent tests this in two calls (toggling SPP at L2) and confirms 
the gain. We verified this MPKI-vs-performance distinction does not 
appear in any prompt file (\texttt{action\_space.md}, \texttt{README.md}); 
the rule emerges from the agent's reasoning, not from retrieval.

\paragraph{Hypothesis-driven ablation}
On mcf seed~3, after observing IPC=0.405 from a configuration with both 
SPP at L2 and SHIP at LLC, the agent voluntarily ablates one component:
\begin{quote}\small\itshape
``To attribute the gain, removing SPP from L2 in call~5: IPC drops to 
0.329. SPP at L2 is responsible. Re-enabling SPP at both L2 and LLC 
in call~6: IPC=0.414, new best.''
\end{quote}
The agent spends one of its budget calls on causal attribution rather 
than on refinement. Black-box optimizers allocate every evaluation to 
fitness improvement; this trace shows the agent reasoning explicitly 
about \emph{why} a design works.

\subsection{A Structural Regularity Across Runs}
\label{sec:priors}

Beyond individual decisions, we observe a consistent structural pattern
across runs. Across our 9 ChampSim runs (3 workloads $\times$ 3 seeds),
8 of 9 best-IPC designs satisfy the L1D~$\leq$~L2C~$\leq$~LLC hierarchy,
and the constraint was \emph{not} stated in \texttt{action\_space.md} at
the time these runs were collected. Uniform random samples from the
unconstrained action space already satisfy this hierarchy roughly 85\%
of the time, so the gap on this small sample is too narrow to support a
strong claim that the agent has internalized the hierarchy as a learned
prior. We report the observation as suggestive: \sysname{}'s outputs
remain consistent with a standard memory-hierarchy convention without
being told to.

\section{Limitations and Future Work}
\label{sec:discussion}

\textbf{LLM Dependence and Efficiency.} A primary limitation is the reliance on high-capacity large language models, which introduces significant computational overhead and API costs during the iterative reasoning process. While this cost is offset by the reduction in simulator evaluations, future work should explore the use of smaller, domain-specific models or distilled student models. Investigating whether specialized ``architectural reasoning'' models can maintain optimization quality while drastically reducing inference latency will be key to scaling the approach.

\textbf{From Optimization to Discovery.} Our current evaluation focuses on design spaces where human designers already possess strong prior heuristics. This allowed us to validate the agent's ability to mirror and accelerate expert-level reasoning. However, the true potential of coding agents lies in exploring complex, non-intuitive design spaces where human intuition is limited, such as novel heterogeneous interconnects or emerging non-von Neumann architectures~\cite{wang2024epim}. Future research will focus on ``zero-shot'' discovery, testing whether \sysname{} can identify novel architectural motifs and generalize new design principles that transcend existing human knowledge.

\textbf{Scope of the Reasoning Ablation.} Our anonymization ablation isolates the contribution of architectural semantics on a single ResNet50 layer; broader coverage across layers, workloads, and domains is a natural next step.

\section{Conclusion}

We present an early but strong empirical case that general-purpose coding agents can serve as practical DSE engines across multiple architecture simulators. By substituting black-box numerical sampling with iterative architectural reasoning, \sysname{} matches or exceeds traditional baselines using up to two orders of magnitude fewer simulator evaluations across DNN accelerator mapping, hardware/software co-design, and CPU cache-hierarchy optimization. 


\bibliographystyle{IEEEtranS}
\bibliography{refs}

\end{document}